\documentclass[a4paper]{jpconf}
\usepackage{graphicx}
\begin{document}
\title{Magnetic models on various topologies}

\author{ F.W.S. Lima$^{1}$ and J. A. Plascak$^{1-3}$}

\address{$^{1}$Dietrich Stauffer Computational Physics Lab, Departamento de F\'{\i}sica,
Universidade Federal do Piau\'{\i} , 64049-550, Teresina, PI, Brazil}
\address{$^{2}$Departamento de F\'{\i}sica,
Universidade Federal de Minas Gerais, C. P. 702, 30123-970, Belo Horizonte, 
MG, Brazil}
\address{$^{3}$Center for Simulational Physics, University of Georgia, 30602 Athens-GA, USA}

\ead{fwslima@gmail.com,pla@hal.physast.uga.edu}

\begin{abstract}
A brief review is given on the study of the thermodynamic properties of spin models 
defined on different topologies like small-world, scale-free networks, 
random graphs 
and regular and random lattices. Ising, Potts and Blume-Capel models 
are considered. They are defined on complex lattices comprising Appolonian, 
Barab\'asi-Albert, Voronoi-Delauny and small-world networks. The main emphasis is given on 
the corresponding phase transitions, transition temperatures, critical
exponents and universality, compared to those obtained by the same models 
on regular Bravais lattices. 
\end{abstract}

\section{Introduction}

The study and characterization of magnetic systems on regular $d$-dimensional
lattices, both experimentally and
theoretically, has been well established during
the last century (see, for instance, the book series edited by Domb and Green,
and also edited by Domb and 
Lebowitz, on {\it Phase Transitions and Critical Phenomena}\cite{domb1,domb2}). Although the transition 
temperatures are model dependent and non-universal, the critical exponents are known to be universal. 
In general, the universality class depends on the spatial lattice dimension $d$,
on the number of components 
$n$ (and symmetry) of the order parameter, and on the range of interactions. The agreement
between theory and experimental results on real compounds has been reported to be excellent in this 
regard.

More recently, there has been a great deal of interest in studying networks, which are different from
the regular crystalline Bravais lattices\cite{bara1,bara2, bara3}. The study of such complex lattices,
also called scale-free networks, has
been mainly motivated by social organizations and computers connectivities\cite{bara2}, among others.
It has been possible to recognize, in this way, networks ranging from networks in nature to networks of
people as well. 

Thus, it turns out quite interesting to understand the behavior of a magnetic system
on such networks and also on random graphs. Despite being theoretical by now,
this will certainly trigger the possibility of synthesizing
experimental realizations of 
materials governed by random-lattices sites. Under this theoretical point of view,
the magnetic model can be defined by usual spin
interactions, where each spin is located on the sites of a complex lattice. The main questions that
arise concern the possible existence of a first- or second-order (critical or
multicritical) 
phase transition, and in the case of a second-order transition,
the corresponding universality class of the model. 

In this brief review we will consider complex lattices 
based on the Voronoi-Delauny (VD) tessellation\cite{vd}, Barab\'asi-Albert (BA)\cite{bara1}, 
Appolonian (AP)\cite{ap}, 
and small-world (SW)\cite{sw} networks, and Erd\"os-R\`enyi (ER)\cite{er} random graphs.
The magnetic models applied to these networks are the Ising, Potts and Blume-Capel models,
among others. In the next section, a short introduction of the Hamiltonian
models and lattices will be presented. In the following sections we will give a
brief summary of what has been done on each network and random graphs and, in
the final
section, we will present a discussion and a summary of the results.

\section{Models, lattices, networks and graphs}

\subsection{Models}

The Ising and Blume-Capel models can be defined by the following Hamiltonian
\begin{equation}
 {\cal H} = -J\sum_{\langle i,j\rangle}\sigma_i\sigma_j - H\sum_{i =
1}^N\sigma_i + D\sum_{i=1}^N\sigma_i^2\;,
\label{ibc}
\end{equation}
where the first sum is over nearest-neighbors on a $d$-dimensional lattice with
$N$ sites, $\sigma_i$ is the state of a spin-$S$ with components $\sigma_i=-S,
-S+1, ... S-1, S$, and $D$ is the crystal field. The general spin-$S$ simple Ising model is
recovered when $D=0$. There is no transition for the one-dimensional model and
in the two and three dimensions the universality class is defined by the
exponents given in Tab. \ref{tabex}.
\begin{table}
\caption{\label{tabex} Order of transition and critical exponents for the Ising
and Potts models
on regular lattices. In two dimensions we have the exact results and in three
dimensions 
the values come from Monte Carlo simulations (see, for instance,
reference\cite{landau}).}
\begin{center}
\begin{tabular}{ccccc} 
\hline \hline 	
model & transition & ~~$\beta/\nu$~~&~~$\gamma/\nu$~~\ &~~$1/\nu$~~\\ 
\hline \hline 
$2d$ Ising-$S$ & 2nd  & $0.125$&$1.75$ & $1$\\
$3d$ Ising-$S$ & 2nd  & $0.5185(15)$&$1.9630(30)$ &$0.630(2)$ \\
$2d$ Potts-$3$ & 2nd  &$2/15=0.133...$&$26/15=1.733...$&$1.2$ \\
$2d$ Potts-$4$ & 2nd  &$0.125$&$1.75$&$1.5$ \\
$2d$ Potts $>4$ & 1st &-&-& -\\
\hline \hline 
\end{tabular}
\end{center}
\end{table}

On the other hand, the $q$-state Potts  model can be written as
\begin{equation}
 {\cal H} = -J\sum_{\langle i,j\rangle}\delta_{\sigma_i\sigma_j}\;,
\label{po}
\end{equation}
where $\delta_{\sigma_i\sigma_j}$ is the Kronecker delta function and now
$\sigma_i=1,2,...,q$. Again, no transition is observed for the one-dimensional
model. In two dimensions, which is the case treated in this work, the universality class
is also given in Tab. \ref{tabex} for $q\le4$ (for $q=2$ it corresponds to the
spin-$1/2$ Ising model). For $q>4$ a first-order
transition takes place.

The above models have also been treated in complex lattices. Some of 
these complex lattices will be briefly described below. 

\subsection{ Directed and undirected Appolonian network}

The Appolonian network is composed of $N=3+(3^{n}-1)/2$ nodes, where $n$
is the generation number and $N$ the node number
\cite{ap,rob1}. On these AP structures we can introduce a disorder, in such a
way that we redirect a fraction $p$
of the links. This redirecting results in a directed network, preserving
the outgoing node of the redirected link
but changing the incoming node. When $p=0$  we have the standard AP networks,
while for $p=1$ we have something similar to random networks \cite{er}. In this
procedure 
of the redirecting links, the number of outgoing links of each node is preserved
even when $p=1$ and
the network still have hubs that  are the most influent nodes. These networks
display a scale-free degree distribution and
a hierarchical structure. In the undirected case there exists the reciprocity of
redirected link.

\subsection{Directed and undirected small-world network}
To generate the directed SW networks \cite{sw} we use a square grid and other
irregular triangular. The disorder
introduced here is identical to the procedure used on AP network cited above
for both cases directed and undirected network.

\subsection{Directed and undirected Erd\"os-R\`enyi random graphs}
An ER random graph is formed by a set of $N$ vertices (sites) connected by $K$
links (bonds)\cite{er} . With  a probability $p$ a given pair of sites is
connected by a bond type $p = 2K/N (N-1)$. The connectivity of a site is defined
as the total
number of bonds connected to it, like $k_{i} =\sum_{j}l_{ij}$, where $l_{ij}= 1$
if there is a link between the sites $i$ and $j$ and $l_{ij}= 0$
otherwise. Random graphs are completely characterized by the mean number of
bonds per site, or the average connectivity $z = p(N-1)$. These links can be
directed or undirected as well.

\subsection{Directed and undirected Barab\'asi-Albert Network}
In the directed BA network, each new site added to the
network selects (a preferential attachment proportional to the number
of previous selections), with connectivity $z$, already existing
sites or nodes as neighbors influencing it; the newly added site does not
influence these neighbors. In the case of the undirected BA network, 
the newly added spin does influence these neighbors.

\subsection{Directed and undirected Voronoi-Delaunay random lattices}

In the undirected VD random lattice the construction of
the lattice obey the following procedure: for each point in a given 
set of points in a plane, we
determine the polygonal cell that contains the region of space
nearest to that point than any other. Two cells are considered
neighbors when they possess an extremity in common (Voronoi tessellation). 
From this Voronoi tessellation, we can obtain the dual lattice,
or triangulation of Delaunay, by the
following procedure.
͑(i) When we have two neighbor cells, a link is placed between 
the two points located in the cells.
͑(ii) From the links, we obtain the triangulation of
space that is called the Delaunay lattice.
͑(iii) The Delaunay lattice is dual to the Voronoi tessellation 
in the sense that its points correspond to cells, links to
edges and triangles to the vertices of the Voronoi tessellation.
The directed VD random lattices is constructed in the same way
as the directed SW network.

\section{Appolonian network}
\subsection{spin-$1/2$ Ising model}

The Ising model has been studied on the triangular Appolonian networks with constant exchange interaction 
$J$ \cite{rob1,rob0} and 
with node-dependent interaction constants $J_{i,j}$ \cite{rob1,rob2}. The corresponding 
thermodynamic and magnetic properties have been obtained by formulating the problem in terms of transfer matrices. 
In both cases
the results show no evidence of any phase transition either for the ferromagnetic or the antiferromagnetic model,
as well as for short or long ranged interactions. Quenched random models, including 
ferromagnetic and antiferromagnetic bond dilution and Ising spin glasses have already been considered \cite{berker}. In this case the ordered phases persist up to infinite temperature over the entire range of disorder.

\subsection{$q$-state Potts model}

The $q$-state Potts model has been studied on Appolonian networks by using Monte Carlo simulations and
transfer matrix techniques\cite{rob3}. In this case, as for the spin-$1/2$ Ising model, no transition has been
detected for any value of $q$.

\section{Erd\"os-R\`enyi random graphs}

\subsection{Ising model}
The spin-$1/2$ and spin-$1$ Ising model have been studied on directed and undirected Erd\"os-R\`enyi random graphs
with different coordination numbers by using Monte Carlo simulations\cite{welER1}. 
For spin-$1/2$, the model exhibits a critical temperature, below which there is a 
spontaneous magnetization, with mean-field exponents. Although the critical temperature depends on being directed 
or undirected graphs, the critical exponents are the same as the mean-field ones for both cases. 
The same qualitative results are also obtained for the spin-$1$ model. However, in this case, one has a first-order
phase transition instead. These results are in agreement with those obtained
for a nonequilibrium model defined
on the same random Erd\"os-R\`enyi lattices \cite{brady,welER2} (in this case,
according to the Grinstein, Jayaprakash and He 
criterion \cite{grinstein}, non-nequilibrium and
equilibrium models
that present up-down symmetry  may belong to the same universality class).

\section{Barab\'asi-Albert Network}

\subsection{Ising model}
The spin-$1/2$ Ising model was first studied on undirected scale-free Barab\'asi-Albert network (UBA)  by
Aleksiejuk {\it et al.} \cite{A_H_S}. Through
Monte Carlo simulations \cite{A_H_S} they have shown that there exists
a Curie temperature which logarithmically increases
with the increasing of the system size $N$. After, Sumour 
{\it et al.} \cite{sumuor1,sumuor2} have treated the Ising model on
a directed Barab\'asi-Albert network (DBA) using standard Glauber kinetic Ising model
to a fixed network. Unlike the previous results presented by Aleksiejuk {\it et al.} \cite{A_H_S},
they have shown that the spin-$1/2$ Ising model on DBA does not exhibit any phase transition for 
temperatures different from zero and confirmed that the corresponding Curie temperature presents an
asymptotic Arrhenius law extrapolation for the relaxation time $\tau$ (defined as the first time when the sign of the
magnetization flips) of the type $1/\ln\tau$ $\propto$ $T$. This means in fact
that at all finite temperatures 
the magnetization eventually vanishes, i.e., no ferromagnetism is present. 
Lima et al. \cite{lima0} have also studied the Ising model with spin S = $1$, $3/2$ and $2$ on DBA and they have shown that no phase transition is present on these DBA networks.

\subsection{$q$-state Potts model}

The Potts model with $q=3$ and $8$ states have been studied on DBA through Monte Carlo simulations 
by Lima \cite{lima1}. Surprisingly, different from the conventional results for the Potts model 
on regular two-dimensional lattices, the Potts model on DBA presented a phase transition of first order
for both values of $q=3$ and $8$.
Sumuor and Lima \cite{lima2} have studied the critical behavior
of Ising and Potts models on semi-directed Barab\'asi-Albert
network (SDBA), recently studied by Sumour and Radwan \cite{sumuor3},
where now the number $N(k)$ of nodes with $k$ links each decays as
$1/k^{\gamma}$ and the exponent $\gamma$ decreases from $3$ to $2$
for increasing $m$, where $m$ is the number of old nodes  which a new node
added to the network selects to be connected with. This behavior is totally 
different from $UBA$ and $DBA$ scale free networks where  $\gamma=3$ is
universal, i.e., independent of $m$. For both Ising and Potts
model the  results showed no
phase transition, in agreement with references\cite{sumuor1,sumuor2}, and a
Curie temperature to diverge at positive
temperatures $T_c(N)$ by a Vogel-Fulcher law, with $T_c(N)$ increasing
logarithmically with network size $N$.

\section{Small-World Network}

\subsection{Ising model}
The one-dimensional spin-$1/2$ Ising model has been studied, via Monte Carlo simulations, 
on small-world network, where the small-world connections have a form proportional to $r_{ij}^{-\alpha}$. It has been
shown that any non-zero value of $\alpha$ destroys the finite-temperature transition in the
thermodynamic limit\cite{jeong}. The results are different when the small-world interactions are
independent of the length ($\alpha=0$), where a finite-temperature phase transition is observed
\cite{git,barrat,peka,hong,kim,mark} with critical exponents smaller than the corresponding ones for the
two-dimensional Ising model.

The two- \cite{zing,mark2} and three-dimensional spin-$1/2$ model have also been treated by Monte Carlo simulations \cite{herr}.
It has been shown that in the thermodynamic limit the phase transition has a mean-field character.

On the other hand, the two-dimensional model has been extended for greater values of spin, namely
spin-$1$, $3/2$ and $2$ on directed small-world networks (DSW) \cite{lima0}. In the DSW network
\cite{sanchez}, Lima {\it et al.} \cite{lima0} showed that the two-dimensional Ising model 
exhibits a second and first-order phase transition for rewiring probability $p = 0.2$ and $0.8$,
respectively, for spin values $S = 1$, $3/2$ and $2$. For values $p > p_c\sim0.25$
this model presents a first-order phase transition\cite{limapla}. These results are, nevertheless, in accordance with the 
those from S\'anchez {\it et al.} \cite{sanchez} for the corresponding non-equilibrium model.

\subsection{$q$-state Potts model}
Recently, Silva {\it et al.} \cite{silva}, through Monte Carlo simulations, have  studied the two-dimensional 
Potts models with $q = 3$ and $4$ states on DSW network. They have found that this model
exhibits a first-order and a
second-order phase transition for $q=3$, depending on the rewiring probability $p$ values. However, 
for $q=4$ the system presents only a
first-order phase transition for any value of $p$ different of zero. This critical behavior on
DSW is different from the Potts model on a regular square lattice, where the
second-order phase transition is present for $q \leq$ $4$ and a first-order phase transition 
takes place for $q > 4$.

\section{Voronoi-Delaunay random lattices} 

\subsection{Ising model}

The  spin-$1/2$ Ising model was first studied, via Monte Carlo simulations, 
on two-dimensional Voronoi-Delaunay random lattices by Espriu {\it et al.} \cite{espriu}.
Their results, obtained with Metropolis update algorithm, showed weak evidence that the critical exponents
belong to the same universality class of the spin-$1/2$ Ising model on regular lattices in two-dimensions.
Janke {\it et al} \cite{janke1,janke2}, using single-cluster Monte Carlo update algorithm \cite{wolf}, 
reweighting techniques \cite{fe} and finite size scaling analysis, also studied the Ising model on 
the two-dimensional Voronoi-Delaunay random lattices and their results confirmed the initial 
evidence obtained by Espriu {\it et al.} \cite{espriu}, namely, that the Ising model on these random lattices 
belong in fact to the same universality class of the Ising model in regular two-dimensional lattices. 
Afterwards, Janke {\it et al} \cite{janke3,janke4}, using the same algorithm earlier applied for 
the two-dimensional Ising model, studied the three-dimensional version on the VD random lattices. 
Again, their results showed that the system
belongs to the same universality class of the Ising model in three-dimensional regular lattices. 
These new results, however, seem to be in disagreement with Harris criterion \cite{harris}. 
On the other hand, Lima {\it et al.} \cite{lima3,lima4} have also studied this model by assuming 
that the exchange coupling $J$ varies with the distance $r$ between
the first neighbors as $J(r)\propto e^{-\alpha r}$, with $\alpha\geq 0$. The model in two and three
dimensions was simulated applying the single-cluster Monte Carlo update algorithm and the 
reweighting technique. The results also showed that this random system belongs
to the same universality class as the pure two and three-dimensional ferromagnetic Ising model.
In addition, within the context of persistence, Lima {\it et al} \cite{lima5} have studied the Ising model
on VD random lattices using the zero-temperature Glauber dynamics.
They showed that the model has exhibited ‘‘blocking’’ effect,
which means that the persistence does not go to zero when time $t\rightarrow \infty$.

Models with spin greater than $1/2$ have also been treated by Monte Carlo simulations. 
While for the spin-$1$ Ising model one also gets the same universality class as for the model on regular
lattices \cite{limapla2}, the same is not true for the spin-$3/2$ case, where the critical exponents
are different\cite{limapla3}.

The spin-$1$ Blume-Capel was treated on directed Voronoi-Delauny lattices\cite{limapla2} and, as in the directed 
small-world case above, a continuous transition has been observed only for rewiring probabilities $p<p_c\sim0.35$,
while for $p>p_c$ a first-order transition takes place.

\subsection{$q$-state Potts model}

Janke {\it et al.} \cite{janke5} have treated the Potts model with $q=8$ states on 
two-dimensional VD random lattices
using the same algorithm and technique from references \cite{janke3,janke4}. They have obtained 
the same first-order behavior as for the model on a regular square lattice. It is worth to stress 
that Chen {\it et al.} \cite{chen} have treated the quenched bond
randomness of this model on regular square lattice and have obtained a phase transition 
which changed from first to second order. Afterwards, Lima {\it et al.} \cite{lima6} have also studied 
the $q=8$ Potts model on VD random lattices using the same
coupling factor $J(r)\propto e^{-\alpha r}$ from references \cite{lima3,lima4}.  
Using Monte Carlo simulations,
and the same procedure from references \cite{janke3,janke4}, they have shown that
for $\alpha = 0$ only first-order transition is present, in agreement with previous works \cite{janke5}.
However, for $\alpha > 0$ the $q=8$ Potts model on VD random lattices
displays a second-order transition with the critical exponents 
$\beta/\nu$ and $\gamma/\nu$ different from the Ising model on regular square lattices. 
For the $q=3$ Potts model\cite{lima7}, the critical exponents are different from those of the model on
a two-dimensional regular lattice, however, the exponents ratio
$\beta/\nu$ and $\gamma/\nu$ seem to be identical to the corresponding ratio on regular lattices.

Recently, Lima \cite{lima8} has studied the $q=4,6$ and $8$ Potts model. 
The $q=4$ model exhibits a second-order transition for $\alpha=0,~0.5$ and $1$ with critical exponents
that are $\alpha$ dependent, and different from those exponents on regular lattices. For $q=6$ and $q=8$,
the system undergoes a first-order transition for $\alpha=0$ and a second-order transition for $\alpha>0$,
with critical exponents again dependent on $\alpha$ and different from those on regular lattices.

\subsection{ Harris criterion and Voronoi-Delaunay random lattices}

The Harris's criterion\cite{harris} for regular lattices 
(for instance, square, triangular, cubic and hyper-cubic lattices) is 
based on the exponent of the 
specific heat of a pure system (free of impurities), to determine 
whether or not the system will change its critical behavior or 
universality class in the presence of impurities or some kind 
of topological disorder. Since then,
it has been well established that when $\alpha < 0$ (not to be 
confused with the exchange decay exponent used in the previous sections) the
critical behavior 
of the specific heat is unchanged when impurities are added 
to the magnetic system in a random way. On the other hand, for $\alpha > 0$
the system with randomness has a critical behavior different 
from the pure system case. However, for the marginal case, i.e,
$\alpha=0$ (two-dimensional Ising model) the Harris's criterion predict that
both
earlier behavior can occur with a pure system in the presence of 
some type of disorder (for the case $\alpha=0$ see, for instance,
\cite{mark3,pla} 
and references therein). However, as mentioned previously, 
Janke {\it et al}\cite{janke1,janke2,janke3,janke4} and Lima {\it et al}
\cite{lima3,lima4} showed that the Harris's criterion fails
on two and three-dimensional Ising model on the VD random lattices.

In order to better understand the Harris's criterion on the VD random 
lattices Janke {\it et al.} \cite{janke6} 
investigated the applicability of this criterion to the case of spin
models on VD random lattices. They were interested in verifying whether the
critical behavior
depends on the degree of spatial correlations present in the random lattices,
which was quantified by a {\it wandering exponent} $w$ . They determined
the numerical value of {\it wandering exponent} $w=0.5$ for
both Ising and Potts models. However,
this value $w=0.5$ corresponds to case $\alpha=0$ of the Harris's criterion
and again, for spin models on VD random lattices, nothing can be
predicted about the corresponding critical behavior of magnetic models.

\begin{table}
\caption{\label{tab} Order of transition and critical exponents (when not
available it is indicated 
by dots) for different models on different random 
lattices. $S$ stands for the value of the Ising spin and $q$ for the number of
states of the Potts model
(or Ising-$S$ and Potts-$q$). $d$ stands for the dimension of the lattice,
and $p$ for the rewiring probability of the corresponding directed random
lattice. 
$\alpha$ gives the range
of the interaction through $J(r)\propto e^{-\alpha r}$ (unless stated
$\alpha=0$).
MF - mean-field. }
\begin{center}
\begin{tabular}{ccccccc} 
\hline \hline \hline	
model & transition & ~~$\beta/\nu$~~&~~$\gamma/\nu$~~\ &~~$1/\nu$~~&
universality& Ref. \\ 
\hline \hline \hline
\multicolumn{7}{c}{Appolonian} \\
\hline \hline
Ising-$1/2$ & no  & -&- & -&-& \cite{rob1}\\
Potts & ~~no~~ &-&-& -&-&\cite{rob2}\\
\hline \hline
\multicolumn{7}{c}{Erd\"os-R\`enyi} \\
\hline \hline \hline
Ising-$1/2$ & 2nd& $1$& $2$& $2$& MF& \cite{welER1,brady,welER2}\\
Ising-$1$& 1st&-&-&-& -&\cite{brady,welER2}\\
\hline \hline 
\multicolumn{7}{c}{Barab\'asi-Albert}\\
\hline \hline \hline
Ising-$S$& no  &-&-& -&-&\cite{sumuor1,sumuor2,lima0}\\
Potts-$q$& no &-&-& -&-&\\
\hline\hline
\multicolumn{7}{c}{Small World}\\
\hline \hline \hline
$1d$ Ising-$1/2$& 2nd & $\sim0.0001$& $0.6(1)$& ...& &\cite{git}-\cite{mark}\\
$2,3d$ Ising-$1/2$& 2nd & $1$& $2$& $2$& MF&\cite{herr}\\
\hline
\multicolumn{7}{c}{ Ising on directed SW}\\
\hline
S-$1/2$ any $p$ & 2nd &$0.44(3)$&$1.129(6)$&$1.14(2)$& ...&\cite{limapla}\\
S-$1,~p<0.25$ & 2nd &$0.40(3)$&$1.22(3)$&$1.16(5)$& ...&\cite{limapla}\\
S-$1,~p>0.25$ & 1st&-&-&-& -&\cite{limapla}\\
\hline
\multicolumn{7}{c}{Potts on directed SW}\\
\hline
$q$=$3$~$p$=$0.1$ & 2nd &$0.24(5)$&$1.5(1)$&...& ...&\cite{silva}\\
$q$=$3$~$p$=$0.9$ & 1st&-&-&-& -&\cite{silva}\\
$q$=$4$ any $p$ & 1st&-&-&-& -&\cite{silva}\\
\hline \hline
\multicolumn{7}{c}{Voronoi-Delauny}\\
\hline \hline \hline
\multicolumn{7}{c}{2d Ising}\\
\hline
S-$1/2$ & 2nd &$0.1208(92)$ &$1.7503(59)$&$0.964(28)$ & Ising&\cite{janke2}\\
S-$1$ & 2nd &$0.135(9)   $&$1.751(4)$  &$1.016(8)$  & Ising&\cite{limapla2}\\
S-$3/2$ & 2nd &$0.331(9)   $&$1.467(9)$  &$1.13(3)$   & ...  &\cite{limapla3}\\
\hline
\multicolumn{7}{c}{3d Ising}\\
\hline
S-$1/2$ & 2nd &$0.51587(82)$&$1.9576(13)$&$1.5875(12)$& Ising&\cite{janke3}\\
\hline
\multicolumn{7}{c}{2d Potts}\\
\hline
$q$=$3$ $\alpha$=$0$  & 2nd &$0.133(13)   $&$1.764(18)    $&$1.19(2)$&
...&\cite{lima7}\\
$q$=$3$ $\alpha$=$0.5$  & 2nd &$0.118(12)   $&$1.751(17)    $&$1.07(1)$&...
&\cite{lima7}\\
$q$=$3$ $\alpha$=$1$  & 2nd &$0.106(12)   $&$1.754(18)    $&$0.94(1)$&...
&\cite{lima7}\\
$q$=$4$ $\alpha$=$0$  & 2nd &$0.143(9)   $&$1.799(6)    $&$1.377(9)$&
...&\cite{lima8}\\
$q$=$4$ $\alpha$=$0.5$  & 2nd &$0.12(2)   $&$1.70(2)    $&$1.13(2)$&...
&\cite{lima8}\\
$q$=$4$ $\alpha$=$1$  & 2nd &$0.10(2)   $&$1.66(4)    $&$0.93(3)$&
...&\cite{lima8}\\
$q$=$6,8$ $\alpha$=$0$   & 1st &-&-&-&- &\cite{janke5,lima6,lima8}\\
$q$=$6$ $\alpha$=$0.5$   & 2nd &$0.122(4)$&$1.53(5)$&$1.09(1)$&...
&\cite{lima8}\\
$q$=$6$ $\alpha$=$1$   & 2nd &$0.14(1)$&$1.56(5)$&$0.91(1)$& &...\cite{lima8}\\
$q$=$8$  $\alpha$=$0.5$ & 2nd &$0.131(7)   $&$1.20(8)    $&$0.83(3)$&
...&\cite{lima8}\\
$q$=$8$  $\alpha$=$1$ & 2nd &$0.126(8)   $&$1.45(6)    $&$0.79(3)$&...
&\cite{lima8}\\
\hline
\multicolumn{7}{c}{ 2d Ising-$1$ on directed VD}\\
\hline
 $p<0.35$ & 2nd &$0.421(7)   $&$1.101(4)$  &$1.105(8)$  &
Ising&\cite{limapla2}\\
 $p>0.35$ & 1st & -         & -        & -       & -    &\cite{limapla2}\\
\hline \hline

\end{tabular}
\end{center}
\end{table}

\section{Discussion}
We should stress that other non-magnetic systems have also been 
studied on such complex
networks, for instance, the majority vote model on Archimedian lattices (see
references\cite{mv1,mv2} and 
references therein) and the non-equilibrium contact-process on Voronoi-Delauny
lattice\cite{dickman}, which
are not being treated here.
In Table \ref{tab} we try to summarize the main results of the magnetic models
treated on random lattices. For
the readers convenience, we have also in Table \ref{tabex} the transition and
critical exponents of the
same models on regular lattices. 

From Table \ref{tab} one can see that there is, in some cases, a change in the
order of the transition and,
in others, no transition at all. When the transition remains continuous, in
general the exponents also
change to another universality class, except for the spin-$1/2$ and spin-$1$
Ising model on the Voronoi-Delauny
lattice and the spin-$1$ Ising model on the directed VD lattice, where the
results for the exponents 
give an indication of belonging to the same 
universality class of the corresponding model on regular latices. It is also
clear that the Harris 
criterion seems not to
be effective for these random lattices, because one cannot say a priori either what the order of the transition 
will be or the universality class of the possible second-order transition. One can argue, of course, 
that the simulations
are still not in the true finite size scaling regime and a change on the results could be obtained by 
considering larger lattices. However, from the quality of the results shown in the simulations it seems not so
plausible such an idea. Thus, apart from some lacking models still to be treated in order to complete Table 
\ref{tab}, the question of the order of the transition and the corresponding universality class of the model
seem to be an open question still to be answered in more general terms

\section{Acknowledgments}
The authors would like to thank M. Novotny, R. F. S. Andrade and N. M. de Araujo for a critical reading of the manuscript. Financial support from CNPq, CAPES, FAPEMIG, and FAPEPI are also gratefully acknowledged.

\end{document}